\documentclass[superscriptaddress,showpacs,aps,twocolumn,prb,floatfix]{revtex4}
\usepackage{eurosym}
\usepackage{graphicx,bm,amsmath,amssymb}
\usepackage[usenames]{color}

\newcommand{\bea}{\begin{eqnarray}}
\newcommand{\eea}{\end{eqnarray}}

\def\beq{\begin{equation}}
\def\eeq{\end{equation}}

\begin{document}

\title{Kondo temperature when the Fermi level is near a step in the
conduction density of states}
\author{J. Fern\'andez}
\affiliation{Centro At\'{o}mico Bariloche and Instituto Balseiro, Comisi\'{o}n Nacional
de Energ\'{\i}a At\'{o}mica, CONICET, 8400 Bariloche, Argentina}
\author{A. A. Aligia}
\affiliation{Centro At\'{o}mico Bariloche and Instituto Balseiro, Comisi\'{o}n Nacional
de Energ\'{\i}a At\'{o}mica, CONICET, 8400 Bariloche, Argentina}
\date{\today }
\author{P. Roura-Bas}
\affiliation{Dpto de F\'{\i}sica, Centro At\'{o}mico Constituyentes, Comisi\'{o}n
Nacional de Energ\'{\i}a At\'{o}mica, CONICET, Buenos Aires, Argentina}
\author{J. A. Andrade}
\affiliation{Centro At\'{o}mico Bariloche and Instituto Balseiro, Comisi\'{o}n Nacional
de Energ\'{\i}a At\'{o}mica, CONICET, 8400 Bariloche, Argentina}

\begin{abstract}
The (111) surface of Cu, Ag and Au is characterized by a band of surface
Shockley states, with constant density of states
beginning slightly below the Fermi energy. These states as well as bulk
states hybridize with magnetic impurities which can be placed above the
surface. We calculate the characteristic low-temperature energy scale, the
Kondo temperature $T_K$ of the impurity Anderson model, as the bottom of the
conduction band $D_s$ crosses the Fermi energy $\epsilon_F$. We find simple
power laws $T_K \simeq |D_s-\epsilon_F|^{\eta}$, where $\eta$ depends on the
sign of $D_s-\epsilon_F$, the ratio between surface and bulk hybridizations
with the impurity $\Delta_s/\Delta_b$ and the ratio between on-site and
Coulomb energy $E_d/U$ in the model.
\end{abstract}

\pacs{72.15.Qm, 73.22.-f, 68.37.Ef}
\maketitle


\section{Introduction}

\label{intro}

The Kondo effect is one of the paradigmatic phenomena in strongly correlated
condensed matter systems \cite{hewson}. It takes place when a localized
magnetic
impurity interacts via an exchange interaction with extended states.
Below a characteristic temperature $T_K$, the impurity spin is screened by
the conduction electrons, and the ground state is a many-body singlet formed
by the impurity spin and the spin of the conduction electrons. Originally
observed in dilute magnetic alloys, the Kondo effect has reappeared more
recently in the context of semiconducting \cite{gold,cro,wiel,grobis,ama}
and molecular \cite{liang,kuba,parks,roch,parks2,serge,vincent} quantum-dot
systems, and in systems of magnetic adatoms (e.g., Co or Mn) deposited on
clean metallic surfaces, where the effect has been clearly observed
experimentally as a narrow Fano-Kondo antiresonance in the differential
conductance ($G(V)=dI/dV$, where $I$ is the current and $V$ the applied
voltage) observed by a scanning tunneling microscope (STM) \cite{li,madha,man,knorr,limot,henzl,serra}.

A STM permits the manipulation of single atoms or molecules on top of a
surface \cite{eigler} and the construction of structures of arbitrary shape
such as quantum corrals.\cite{crom,heller,man} The differential conductance
measured by the STM is in general proportional to the local density of metal
states, and it has contributions from bulk and surface states.\cite{fiete,revi} 
These contributions are weighted differently by the STM tip due
to the different decay rate of the wave functions out of the surface.\cite{plihal} 
The effect of the different distance dependence of tunneling
processes involving $3d$ and $s/p$ states has been observed recently for Fe$_2$N on Cu(001).\cite{taka}

The (111) surfaces of Cu, Ag and Au were used as the substrate for many
observations of the Kondo effect \cite{li,madha,man,knorr,limot,henzl,serra,zhao,kome,mina,iancu} 
and have the property that a
parabolic band of two-dimensional Shockley surface states, confined to the
last few atomic planes exists.\cite{hulb,cerce,abd} This band is uncoupled
to bulk states for small wave vectors parallel to the surface, due to the
presence of a bulk-projected band gap at the center of the surface Brillouin
zone.\cite{hulb} These surface states represent an almost ideal example of a
two-dimensional electron gas on a metal surface. The effective mass is
between 0.31 and 0.38 of the electron mass,\cite{crom,euc,limot} and the
constant surface density of states begins at a step which lies below the
Fermi energy by an energy $\approx 450$ meV for Cu,\cite{knorr} $\approx 475$
meV for Au,\cite{cerce} and $\approx 67$ meV for Ag.\cite{limot} The
corresponding steps have been observed in STM experiments.\cite{cerce,limot}
Interestingly, it has been shown recently that the Shockley surface states 
can be thought as topologically derived surface states from a topological 
energy gap lying about $\sim$ 3 eV above the Fermi energy.\cite{claudia}

The surface states are expected to be more sensitive to adatoms at the
surface, and this fact has been used to confine electrons in corrals or
resonators built from different adatoms.\cite{heller} Recently, it has been
shown that the effect on the surface density of states of resonators of Co
and Ag adatoms built on the Ag(111) observed by an STM can be modeled by
the effect of an attractive potential at the position of the adatoms on free
electrons in two dimensions.\cite{joaq} In a famous experiment, a Co
atom acting as a magnetic impurity was placed at one focus of an elliptical
quantum corral built on the Cu(111) surface. A Fano-Kondo antiresonance was
observed in the differential conductance not only at that position, but also
with reduced intensity at the other focus.\cite{man} This \textquotedblleft
mirage\textquotedblright\ can be understood as the result of quantum
interference in the way in which the Kondo effect is transmitted from one
focus to the other by the different eigenstates of surface conduction
electrons inside a hard-wall ellipse.\cite{revi,rap,por,wei,hal,ali2,wil}
This experiment reveals that conducting surface states have an important
hybridization with the impurity, although other experiments suggests that
the hybridization of bulk electrons plays the dominant role in this effect.
\cite{knorr,limot,henzl}
Interestingly, a Kondo resonance with $T_K \sim 180$ K was obtained
for a system of a molecule containing a magnetic Co atom on a Si substrate prepared in such a way 
to have a surface metallic state on top of insulating Si.\cite{li2} 
In this case clearly bulk states do not contribute to the observed Kondo resonance.
Some calculations suggest that surface states give an important contribution to the
Fano-Kondo antiresonance,\cite{merino} while others obtain a contribution of 1/36 or less
depending on the orbital.\cite{barral}
A recent study for Co on Cu(111) 
suggests that the line shape of the Kondo resonance is affected by the presence of surface states.\cite{Baruselli} 
From the mirage intensity it has been estimated that the
coupling to the surface states is at least 1/10 that of the bulk.\cite{revi}
Recent experiments for a Co impurity on Ag(111) \cite{moro} in which the 
surface density of states at the Fermi level $\rho_s(0)$ has been modified by means
of resonators \cite{moro,joaq} obtain an increase of a factor larger than 2 in $T_K$
as a consequence of a moderate increase in $\rho_s(0)$, indicating 
a very important contribution of the surface states to the Kondo effect. 

As we show in the next section, for hybridization independent of the energy
(as usually assumed to be a reasonable approximation), the presence of
a step in the conduction spectral density near the Fermi energy has dramatic effects
on the Kondo temperature $T_K$ (the characteristic energy scale of the Kondo effect).
It has been shown that the bottom of the surface band $D_{s}$
can be changed by alloying the different noble metals at the surface.\cite{cerce,abd}
In fact this displacement has been measured by STM,\cite{cerce} and 
it should be possible to use the STM to measure also the Fano-Kondo antiresonance
and determine $T_K$. One may suspect that disorder affects the sharp onset 
of the surface states at $D_s$ but the states at this onset have very long
wave length averaging the disorder. Furthermore, experiments on epitaxial Ag(111) 
films on Si(111)-($7 \times 7$)
have shown that it is possible to change $D_s$ and make it cross the Fermi energy  
by strain.\cite{neu} We note that recent developments in scientific
instruments using a piezoelectric vice,\cite{andy1} have been used to obtain both
uniaxial compression and uniaxial tension on different samples. As an example, both effects  
increase the superconducting critical temperature of Sr$_2$RuO$_4$.\cite{andy2}

In this work we calculate the Kondo temperature as a function of the bottom
of the surface band, using different techniques: poor man's
scaling (PMS)\cite{hewson,and} on the effective Kondo model, non-crossing
approximation (NCA),\cite{hewson,bickers}, slave bosons in the mean-field
approximation (SBMFA),\cite{hewson,cole,newns} and  numerical-renormalization group 
(NRG).\cite{hewson,Bulla,Wilson,Cragg,Krishna,Campo} 
These approaches are known
to reproduce correctly the relevant energy scale $T_{K}$ and its dependence
of parameters in cases in which the conduction density of states is smooth. 
As we shall see, the presence of the surface states introduces some
complications, due to the divergence of the one-body part of the self energy
[Eq. (\ref{ss}) below] at energies near the step, but this can be handled
by the NCA, which exactly incorporates arbitrary conduction
densities and hybridizations in its integral equations.\cite{kroha} 

The PMS is a perturbative
approach that integrates out progressively a small portion of the conduction
states lying at the bottom and at the top of the conduction bands,
renormalizing the Kondo coupling $J$. 
It ceases to be valid when $|D_{s}-\epsilon _{F}|\sim J$, 
where $\epsilon _{F}$ is the Fermi energy.\cite{hewson,and} 

The NCA is equivalent to a sum of an infinite series of
diagrams in perturbations in the hybridization.\cite{hewson,bickers} In
contrast to NRG in which finite-energy features
are artificially broadened due to the logarithmic discretization of the
conducting band \cite{zitko,loig}, NCA correctly describes these features. For
instance, the intensity and the width of the charge-transfer peak of the
spectral density (the one near the dot level $E_{d}$) was found \cite{capa}
in agreement with other theoretical methods \cite{pru,logan} and experiment 
\cite{haug}. Furthermore, it has a natural extension to non-equilibrium
conditions \cite{wingreen} and it is specially suitable for describing
satellite peaks of the Kondo resonance, as those observed in Ce systems,\cite{reinert,ehm} 
or away from zero bias voltage in non-equilibrium transport.\cite{tosi_2,NFL,st} 
An alternative to NCA for non-equilibrium problems is renormalized perturbation 
theory, but it is limited to small bias voltage.\cite{hbo,ng}

The
SBMFA, as the NCA uses a pseudoparticle representation, but in contrast to
the latter, neglects the dynamics of the pseudoboson and takes it in
average.\cite{hewson,cole,newns} In spite of this and the fact that the
charge-transfer peak is lost, the spectral density near the Fermi level and
low-energy properties are well described. For this reason it has been
successful in describing several Kondo systems,\cite{rome,si,mole} including
adatoms or molecules on the (111) surface of Cu or noble 
metals.\cite{revi,rome,mole,joa2}

The NRG is a very accurate technique that has been used for many 
problems. \cite{Baruselli,Bulla,Wilson,Cragg,Krishna,Campo}
However, as stated above, in some cases it misses some finite energy features.
This fact seems to introduce some difficulties in our problem, as we shall show.

In Sec. \ref{model}, we describe the impurity Anderson model, the
particularities of our case, and its Kondo limit used in PMS. The results
are presented in Sec. \ref{res} and Sec. \ref{sum} contains a summary and
discussion.

\section{Model and formalism}

\label{model}

\subsection{Hamiltonian}

The Hamiltonian can be written as

\begin{eqnarray}
H &=&\sum_{k\sigma }\varepsilon _{k}^{s}s_{k\sigma }^{\dagger }s_{k\sigma
}+\sum_{k\sigma }\varepsilon _{k}^{b}b_{k\sigma }^{\dagger }b_{k\sigma
}+E_{d}\sum_{\sigma }d_{\sigma }^{\dagger }d_{\sigma }+  \notag \\
&&+U\sum d_{\uparrow }^{\dagger }d_{\uparrow }d_{\downarrow }^{\dagger
}d_{\downarrow }+\sum_{k\sigma }V_{k}^{s}[d_{\sigma }^{\dagger }s_{k\sigma }+%
\text{H.c.}]+  \notag \\
&&+\sum_{k\sigma }V_{k}^{b}[d_{\sigma }^{\dagger }b_{k\sigma }+\text{H.c.}].
\label{ham}
\end{eqnarray}%
where $d_{\sigma }^{\dagger }$ creates an electron with spin $\sigma $ at
the relevant orbital of the magnetic impurity (assumed non-degenerate) and $%
s_{k\sigma }^{\dagger }$ ($b_{k\sigma }^{\dagger }$) are creation operators
for an electron in the $k^{th}$ surface (bulk) conduction eigenstate.

The spectral density of electrons at the magnetic impurity is

\begin{equation}
\rho _{d\sigma }(\omega )=\frac{1}{2\pi i}[G_{d\sigma }(\omega -i\epsilon
)-G_{d\sigma }(\omega +i\epsilon )],  \label{ro}
\end{equation}%
where $\epsilon $ is a positive infinitesimal. Calling $z=\omega +i\epsilon $
($z=\omega -i\epsilon $), the retarded (advanced) Green's function at the
interacting QD can be written in the form \cite{lan,mir}

\begin{equation}
G_{d\sigma }(z)=\frac{1}{z-E_{d}-\Sigma _{0\sigma }(z)-\Sigma _{d\sigma }(z)}%
,  \label{gd}
\end{equation}%
where $\Sigma _{d\sigma }(z)$ is the self-energy due to the interaction $U$
and $\Sigma _{0}(z)$, the non-interacting part of the self-energy (present
also for $U=0$) is

\begin{eqnarray}
\Sigma _{0\sigma }(z) &=&\Sigma _{0\sigma }^{b}(z)+\Sigma _{0\sigma }^{s}(z),
\notag \\
\Sigma _{0\sigma }^{c}(z) &=&\sum_{k}\frac{|V_{k}^{c}|^{2}}{z-\varepsilon
_{k}^{c}},  \label{s0}
\end{eqnarray}%
where $c=b$ or $c=s$. As usual in this type of problems, in which the bulk
contribution has no special features near the Fermi level, we assume a
constant density of bulk states $\rho _{b}$ extending in a wide range from $%
-D$ to $D$, and a constant hybridization $V_{k}^{b}$. Defining $\Delta
_{b}=\pi \rho _{b}|V_{k}^{b}|^{2}$, for energies near the Fermi level $%
\epsilon _{F}$ ($D\gg |\omega -\epsilon _{F}|$), we can neglect the real
part of the bulk contribution to $\Sigma _{0\sigma }(z)$ \cite{revi} and it
becomes simply

\begin{equation}
\Sigma _{0\sigma }^{b}(\omega +i\epsilon )=-i\Delta _{b}.  \label{sb}
\end{equation}%
The surface contribution do the density of states $\rho _{s}$ is constant
and begins near the Fermi energy at $D_{s}$. For simplicity we assume that
it ends also at $D$ as the bulk one. Then

\begin{equation}
\Sigma _{0\sigma }^{s}(\omega +i\epsilon )=-\frac{\Delta _{s}}{\pi }\ln
\left( \frac{D-\omega }{|D_{s}-\omega |}\right) -i\Delta _{s}\theta (\omega
-D_{s}),  \label{ss}
\end{equation}%
where $\Delta _{s}=\pi \rho _{s}|V_{k}^{s}|^{2}$ and $\theta (\omega )$ is
the step function. Thus, the non-interacting ($U=0$) Green's function can be
written in the form

\begin{equation}
G_{d\sigma }^{0}(z)=\frac{1}{z-\tilde{E}_{d}(\omega )+i[\Delta _{b}+\Delta
_{s}\theta (\omega -D_{s})]},  \label{gd0}
\end{equation}%
where

\begin{equation}
\tilde{E}_{d}(\omega )=E_{d}-\frac{\Delta _{s}}{\pi }\ln \left( \frac{%
D-\omega }{|D_{s}-\omega |}\right)   \label{edef}
\end{equation}%
is an effective energy of the localized level.

\subsection{The Kondo limit}

The Kondo effect takes place for dot occupations near one. This condition in
terms of the parameters means $\Delta _{b}+\Delta _{s}\theta (\epsilon
_{F}-D_{s})\ll \epsilon _{F}-\tilde{E}_{d}(\epsilon _{F})$, $\tilde{E}%
_{d}(\epsilon _{F})+U-\epsilon _{F}$. The Kondo Hamiltonian is obtained from
the Anderson one by means of a canonical transformation to second order in
the hybridization $V_{k}^{b}$ and $V_{k}^{s}$.\cite{hewson,sw} To simplify
the problem, using the fact that all physical quantities depend on conduction spectral
densities and hybridization only through the products $\Delta _{b}$ and $\Delta
_{s}$, we consider an equivalent problem in which both hybridizations are
equal to the bulk one and the new surface density of states $\tilde{\rho}%
_{s} $ is modified accordingly in such a way that

\begin{equation}
\Delta _{s}=\pi \tilde{\rho}_{s}|V_{k}^{b}|^{2}.  \label{renor}
\end{equation}%
Then, the effective Kondo interaction can be written in the form

\begin{equation}
H_{K}=J\sum\limits_{kq}[S^{+}c_{k\downarrow }^{\dagger }c_{q\uparrow
}+S^{-}c_{k\uparrow }^{\dagger }c_{q\downarrow }+S_{z}(c_{k\uparrow
}^{\dagger }c_{q\uparrow }-c_{k\downarrow }^{\dagger }c_{q\downarrow })],
\label{hk}
\end{equation}%
where $c_{i\sigma }^{\dagger }$ ($i=k,q$) includes both, bulk and surface conduction
electrons, and the spin operators act on the localized spin ($|\sigma
\rangle =d_{\sigma }^{\dagger }|0\rangle $). The interaction is calculated
for conduction states near the Fermi energy $\epsilon _{F}$ and becomes for $%
V_{k}^{b}=V$ near $\epsilon _{F}$

\begin{equation}
J=\frac{|V|^{2}}{\epsilon _{F}-\tilde{E}_{d}(\epsilon _{F})}+\frac{|V|^{2}}{%
\tilde{E}_{d}(\epsilon _{F})+U-\epsilon _{F}}.  \label{j}
\end{equation}

\section{Results}
\label{res}

In this section we present the results of the different techniques used. For
simplicity, from now on we choose the origin of energies at $\epsilon _{F}=0$%
.

\subsection{Poor man's scaling}
\label{pmsr}

Here used the PMS \cite{hewson,and}  for the Kondo Hamiltonian, which allows
us to obtain analytical results in the Kondo regime and for $|D_{s}|\gtrsim J$. 
The idea is very simple. Integrating out successively the states on the
top and bottom of the conduction band renormalizing $J$, one has the same
problem but with a smaller total band width (from $-D^{\prime }$ to $%
D^{\prime }$ and larger Kondo interaction $J(D^{\prime })$. Proceeding in
this way until $D^{\prime }=|D_{s}|$, the step in the density of states
disappears and one recovers the ordinary Kondo problem. In its simplest form
(to second order in $J$), the equation for the change in $J$ in the range where
the running cutoff $D'$ is larger than $|D_s|$ is

\begin{equation}
\frac{dJ}{d \ln D^\prime} = -(2\rho _{b}+\tilde{\rho}_{s})J^{2}.  \label{dj}
\end{equation}
The different factors in front of the densities is due to the fact that the
surface part only acts at the top of the band.

Integrating Eq. (\ref{dj}) one has an equation for $J(|D_{s}|)$

\begin{equation}
D\exp \left[ -\frac{1}{(2\rho _{b}+\tilde{\rho}_{s})J}\right] =|D_{s}|\exp %
\left[ -\frac{1}{(2\rho _{b}+\tilde{\rho}_{s})J(|D_{s}|)}\right] ,
\label{jds}
\end{equation}%
and now one can use the expression for the Kondo temperature for a band with 
$|D_{s}|$, Kondo interaction $J(|D_{s}|)$ and density $\rho =\rho _{b}+%
\tilde{\rho}_{s}$ ($\rho =\rho _{b}$) for $D_{s}<0$ ($D_{s}>0$):

\begin{equation}
T_{K}\simeq |D_{s}|\exp \left[ -\frac{1}{2\rho J(|D_{s}|)}\right] .
\label{tk}
\end{equation}%
Using Eqs. (\ref{renor}), (\ref{jds}) and (\ref{tk}) one obtains for $D_{s}<0
$

\begin{eqnarray}
T_{K} &\simeq &|D_{s}|^{\nu }D^{1-\nu }\exp \left[ -\frac{1}{2J(\rho _{b}+%
\tilde{\rho}_{s})}\right] ,  \notag \\
\nu  &=&\frac{\Delta _{s}}{2(\Delta _{b}+\Delta _{s})},  \label{tkn}
\end{eqnarray}%
and for $D_{s}>0$

\begin{eqnarray}
T_{K} &\simeq &D_{s}^{-\mu }D^{1+\mu }\exp \left[ -\frac{1}{2J\rho _{b}}%
\right] ,  \notag \\
\mu  &=&\frac{\Delta _{s}}{2\Delta _{b}}.  \label{tkp}
\end{eqnarray}%
From Eqs. (\ref{edef}) and (\ref{j}) one has

\begin{eqnarray}
\frac{1}{J\rho _{b}} &=&\frac{\pi }{\Delta _{b}U}%
[-E_{d}(E_{d}+U)-(2E_{d}+U)x-x^{2}],  \notag \\
x &=&\frac{\Delta _{s}}{\pi }\ln \frac{|D_{s}|}{D}.  \label{jef}
\end{eqnarray}%
If $|D_{s}|$ is not too small, one can neglect the term in $x^{2}$ in
comparison with the first term in square brackets in Eq. (\ref{jef}). In any
case for small $|D_{s}|$, the PMS ceases to be valid. Using this
approximation and replacing Eq. (\ref{jef}) in Eqs. (\ref{tkn}) and 
(\ref{tkp}) we obtain

\begin{eqnarray}
T_{K} &\simeq &A|D_{s}|^{\eta }D^{1-\eta }\exp \left[ \frac{\pi
E_{d}(E_{d}+U)}{2U(\Delta _{b}+\Delta _{s})}\right] ,  \notag \\
\eta  &=&\frac{\Delta _{s}}{(\Delta _{b}+\Delta _{s})}(1+\frac{E_{d}}{U}),%
\text{ if }D_{s}<0.  \notag \\
T_{K} &\simeq &BD_{s}^{\zeta }D^{1-\zeta }\exp \left[ \frac{\pi
E_{d}(E_{d}+U)}{2U\Delta _{b}}\right] ,  \notag \\
\zeta  &=&\frac{\Delta _{s}E_{d}}{\Delta _{b}U},\text{ if }D_{s}>0,
\label{tkf}
\end{eqnarray}%
where to second order in $J$, $A=B=1$. This is the main result of this
section. As expected, the expressions are correct in the obvious limits, 
$\Delta _{s}=0$, $D_{s}=-D$, and $D_{s}=D$, although the prefactor is
somewhat larger that a more accurate one that can be obtained including
terms up to third order in $J$ in PMS.\cite{hewson,solyom} Including these
terms in the present case is more involved than in the usual case in which
an electron-hole symmetric conduction band is assumed. Terms of order 
$J^{3}/(D^{\prime }+|D_{s}|)$ appear when calculating $dJ/dD^{\prime }$ and
an analytical solution of the differential equation is not possible. When 
$|D_{s}|=D$, the band recovers the electron-hole symmetry and we can borrow
previous results, that we display for later use:

\begin{eqnarray}
A &=&\sqrt{2\rho _{b}J(1+\Delta _{s}/\Delta _{b})},  \notag \\
B &=&\sqrt{2\rho _{b}J},  \label{ab}
\end{eqnarray}%
where $\rho _{b}J$ is given by Eq. (\ref{jef}).

For comparison with the results of other techniques, we note the limiting
values of the exponents for infinite $U$ depending if either $E_{d}$ or 
$E_{d}+U$ remains finite

\begin{eqnarray}
\eta  &=&\frac{\Delta _{s}}{(\Delta _{b}+\Delta _{s})}\text{, }\zeta =0 
\notag \\
\text{if }U &\rightarrow &+\infty ,E_{d}\text{ finite,}  \label{expn}
\end{eqnarray}

\begin{eqnarray}
\eta &=&0\text{, }\zeta =-\frac{\Delta _{s}}{\Delta _{b}}  \notag \\
\text{if }U &\rightarrow &+\infty ,E_{d}+U\text{ finite.}  \label{expp}
\end{eqnarray}

\subsection{Non-crossing and slave-boson mean-field approximations}
\label{ncab}

The non-crossing approximation (NCA) is a diagrammatic technique that
reproduces correctly the Kondo temperature of the spin-1/2 impurity Anderson model
in the limit $U \rightarrow +\infty$.\cite{hewson,bickers} Unfortunately, this is 
not the case for finite $U$ \cite{pruschke89, haule01, tosi11}. 
Therefore, we restrict the NCA calculations to $U \rightarrow +\infty$.
To determine the value of the Kondo temperature $T_K$, we calculate
the conductance through the magnetic impurity as a function of temperature $G(T)$ and 
look for the temperature such that $G(T_K)=G_0/2$, where $G_0$ is the ideal 
conductance of the system (reached for $T=0$ and occupancy 1 of the dot level).
Alternative definitions of $T_K$ differ in factor of the order of 1,\cite{tositk}
which is not relevant to us, since we are interested in the dependence of $T_K$ with $D_s$.

\begin{figure}[h]
\includegraphics[clip,width=8cm]{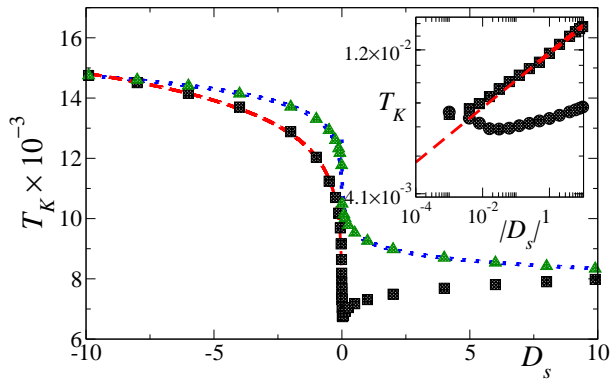}
\caption{(Color online) Kondo temperature as a function of the bottom of
the conduction band. Squares and circles: NCA. Triangles: SBMFA. Dashed (dot) line: 
fit to the NCA (SBMFA) results. The inset shows the NCA results in log-log scale.
Parameters are $E_d=-4$, $\Delta_b=1$, $\Delta_s=0.1$.}
\label{p1-1}
\end{figure}

In the following we take $\Delta_b + \Delta_s$ of the order of the unit of energy and set 
$D=10$. We begin taking $E_d=-4$ so that the system is in the Kondo regime.
In Fig. \ref{p1-1} we show the resulting $T_K$ as a function of $D_s$ for 
a ratio $\Delta_s/\Delta_b=1/10$, the lower limit estimated on the basis of
the mirage experiment for a Co impurity inside an elliptical corral on the Cu(111) 
surface.\cite{revi} A fit of the NCA results with the function $T_K=C|D_s|^{\eta}$  
in the interval $-10 \leq D_s \leq -0.5$ gives $C=0.01204$ and $\eta=0.09039$.
The exponent is in very good agreement with the PMS result $\eta=1/11=0.09091$ from 
Eq. (\ref{expn}). For smaller $|D_s|$, in particular when $|D_s| \sim T_K$, the PMS ceases 
to be valid. The NCA gives a continuous function for $T_K$ with a finite value for $D_s=0$.
For positive $D_s$, PMS predicts a constant $T_K$ [Eqs. (\ref{tkf}) and (\ref{expn})].
As a first approximation, the NCA results are consistent with this. However, for small 
$D_s$, $T_K$ decreases by about 17\%. There is also a non-monotonic behavior with a minimum near 
$D_s=0$. Concerning the magnitude of $T_K$, the NCA value for $D_s=-D$ is 0.0148, while
Eq. (\ref{tkf}) with $A=0.42$ given by Eq. (\ref{ab}) gives $T_K=0.0139$ in good agreement with 
the NCA result. For $D_s=D$, the corresponding values are 0.0080 and 0.0075 respectively.
The NCA values are near 7\% higher.

In Fig. \ref{p1-1} we also show the result of $T_K$ using the SBMFA
for the same parameters. In this case, 
we define $T_K$ as the half width at half maximum of the Kondo resonance in the spectral 
density of states. The results shown in the figure were multiplied by 0.447 
so that they coincide
with those of the NCA for $D_s=-D$. Curiously, this factor is similar to $A=0.42$ discussed above.
Although the SBMFA gives the correct order of magnitude
of $T_K$ for not too small $D_s$, the dependence with $D_s$ is not reproduced, 
although the function can still be fit with power laws.
For negative (positive) $D_s$ the fit gives an exponent 0.0442 (-0.045). 
Curiously, these values are close to the values $\nu=1/22=0.0454$ and $-\mu=-1/20=-0.05$ given
in Eqs. (\ref{tkn}) and (\ref{tkp}). This fact suggest that the SBMFA
misses the renormalization of $E_d$ due to the step in the density of states [Eq. (\ref{edef})].
This is a shortcoming of the approximation for small $|D_s|$, and might be a problem for Ag, for 
which $-D_s$ is only 67 meV. However, for Co on Ag(111), for example, the ratio $T_K/|D_s|$
is still slightly below 0.1.\cite{serra}
For $T_K/|D_s|=0.1$ the ratio in Fig. 1 between the SBMFA and NCA
is 1.11. As we shall see this ratio increases with $\Delta_s/\Delta_b$ but remains below an order 
of magnitude.

\begin{figure}[h]
\includegraphics[clip,width=8cm]{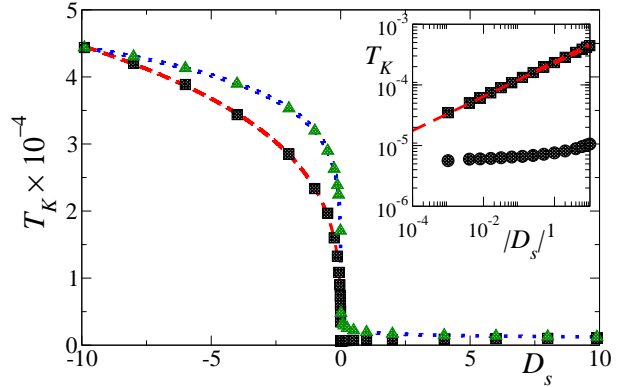}
\caption{(Color online) Same as Fig. \ref{p1-1} for  $\Delta_b=0.5$ and $\Delta_s=0.2$.}
\label{p2p5}
\end{figure}

In Fig. \ref{p2p5} we show  $T_K$ vs $D_s$ for a larger ratio $\Delta_s/\Delta_b=2/5$.
The first obvious change with respect of the previous case is that now $T_K$ for negative 
and large in magnitude $D_s$ is about 50 times larger than for positive and large $D_s$, 
while in the previous case this factor was near 2. This is due to the fact that for negative
$D_s$ both surface and bulk states contribute to the Kondo effect while for positive $D_s$
only the bulk states remain at the Fermi energy. Fitting as before the NCA results for 
$D_s < -0.125$ by a power law we obtain an exponent $\eta=0.2772$, again very near the PMS 
value $\eta=2/7=0.2857$. For positive $D_s$, $T_K$ increases slightly. 
In this case for $D_s=-D$, Eqs. (\ref{tkf}) and Eq. (\ref{ab}) give $A=0.33$ and 
$T_K=4.16 \times 10^{-3}$ while the NCA value is $4.44 \times 10^{-3}$.
For $D_s=D$ the corresponding values are $9.8 \times 10^{-6}$ and $1.06 \times 10^{-5}$.
Again the NCA values are near or 8\% larger than the PMS results. 

The fitting of the SBMFA results for $|D_s| > 0.125$ gives an exponent 0.142 for
$D_s<0$ again near to $\nu=1/7=0.143$ and -0.199 for $D_s > 0$ very near to $-\mu=-1/5$. 
For the comparison in Fig. \ref{p2p5}, the SBMFA results were multiplied by 0.35, near to 
$A=0.33$, suggesting as before that SBMFA gives a value near to the PMS result to second order in 
$J$, while the NCA seems to capture higher order corrections. An additional factor
3.16 exists between SBMFA and NCA results for the point where for the NCA $T_K/|D_s|=0.1$.

\begin{figure}[h]
\includegraphics[clip,width=8cm]{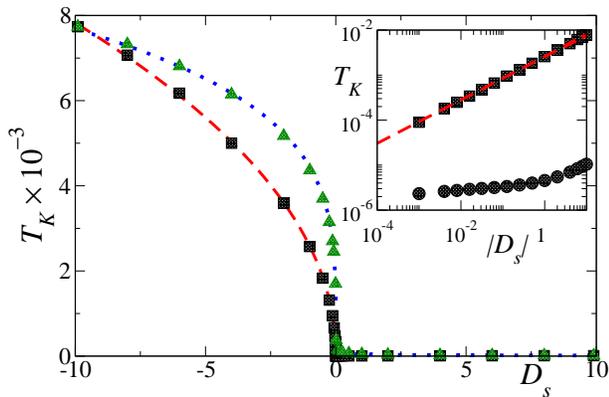}
\caption{(Color online) Same as Fig. \ref{p1-1} for  $\Delta_b=0.5$ and $\Delta_s=0.5$.}
\label{p5p5}
\end{figure}

The case $\Delta_s=\Delta_b$ is displayed in Fig. \ref{p5p5}. The general features are similar to those of the 
previous two figures, with a more dramatic difference in $T_K$ between negative and positive
$D_s$ reaching three orders of magnitude. The NCA exponent of the fit for $D_s < -0.01$ gives
$\eta=2/7=0.482$ near to the expected PMS value 1/2. As above, $T_K$ increases slowly for $D_s >0$.
Here for $D_s=-D$, the PMS results give $A=0.40$ and 
$T_K=7.48 \times 10^{-3}$ while the NCA value is $7.74 \times 10^{-3}$ (3\% larger).
For $D_s=D$ the PMS value is the same as in the previous case and the NCA value is 
$1.05 \times 10^{-5}$ (7\% larger). 

Fitting of the SBMFA results for $|D_s| > 0.01$ gives exponents 0.239 and -0.496, 
near to the expected values 1/4 and -1/2 according to the analysis of the previous figures.
The SBMFA results in the figure were were multiplied by 0.412. To see the difference between NCA
and SBMFA results for small $D_s$, we have calculated the ratio between both of them 
when for the NCA $T_K/|D_s|=0.1$. Here the SBMFA gives a value 5.54 larger in the figure,
or 13.4 times larger taken into account both factors.

The previous results were taken for infinite $U$ and finite $E_d$ for which the occupancy
of the magnetic impurity fluctuates between 0 and 1, although it is near 1 in the Kondo limit.
Another possible way to take this limit is to send $E_d \rightarrow -\infty$ together with
$U \rightarrow +\infty$ keeping $E_d+U$ constant. This case correspond to fluctuations between a singly 
and double occupied impurity. The expected exponents are
given by Eqs. (\ref{expp}). The behavior is qualitatively different from that studied so far in that
$T_K$ is expected to be constant for $D_s <0$ and divergent for $D_s >0$ and not too small $D_s$.
To solve one of these cases with the NCA we have performed a special electron-hole transformation that
reflects the conduction bands around the Fermi energy and $E_d$ is changed to $-E_d -U$.\cite{loig}.

\begin{figure}[h]
\includegraphics[clip,width=8cm]{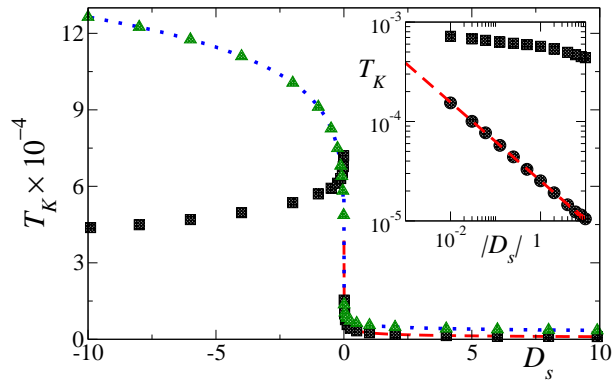}
\caption{(Color online) Same as Fig. \ref{p1-1} for $E_d+U=4$, $\Delta_b=0.5$ and $\Delta_s=0.2$.}
\label{hp2p5}
\end{figure}

The results in the original electron representation for $E_d+U=4$ and an intermediate
ratio $\Delta_s/\Delta_b=2/5$ are shown in Fig. \ref{hp2p5}. In contrast to the previous cases, for 
negative $D_s$, $T_K$ increases with increasing $D_s$ reaching near 50\% for $D_s=0$. 
According to Eqs. (\ref{expp}) one would expect a constant behavior for $D_s<0$. 
The difference might be due 
to terms of higher order in $J$ not included in our PMS treatment. Instead, the SBMFA predicts a
decreasing $T_K$ with increasing $D_s$, which is not expected. For positive $D_s$ a fit of the NCA data gives 
an exponent $\zeta=-0.395$ in very nice agreement with $\zeta=-0.4$ given by Eqs. (\ref{expp}). 
Fitting of the SBMFA results gives exponents 0.142 for negative $D_s$ and -0.200 for positive 
$D_s$, again near to the values $\nu=1/7$ and $-\mu=1/5$ and in disagreement with NCA and PMS.

The PMS results for $T_K$ at $|D_s|=D$ are the same as for the case of Fig. (\ref{p2p5}) because of
electron-hole symmetry in the absence of the step, namely $T_K=4.16 \times 10^{-3}$ and 
$9.8 \times 10^{-6}$, while the NCA values are very similar to those of that case $4.38 \times 10^{-3}$ and  
$1.04 \times 10^{-5}$. The difference might be due to an error of the order of 1\% in determining $T_K$

\subsubsection{Spectral density within the NCA}
\label{rho}

\begin{figure}[h]
\includegraphics[clip,width=8cm]{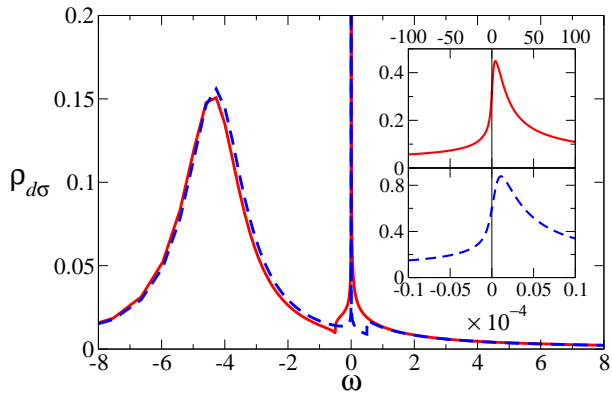}
\caption{(Color online) Spectral density of states within NCA for $E_d=-4$, $\Delta_b=\Delta_s=0.5$ and 
two values of $D_s$,
full line: $D_s=-0.5$, dashed line: $D_s=0.5$. The insets show details of the peaks near $\omega=0$.}
\label{electro}
\end{figure}

The spectral density at the magnetic impurity for each spin 
$\rho _{d\sigma}(\omega )$ is given by Eq. (\ref{ro}). Due to the structure of the
non-interacting part of the self-energy $\Sigma _{0\sigma }$ (see Section \ref{model}) 
one expects that some anomaly might be present in  $\rho_{d\sigma }(\omega )$ for  
$\omega \sim D_{s}$, particularly for large $\Delta _{s}$.
In Fig. \ref{electro} we show $\rho _{d\sigma}(\omega )$ calculated with NCA for two small values 
of $D_s$ and other parameters as in Fig. \ref{p5p5}. The main difference with 
usual NCA results for the spectral density is that the step in the conduction density of
states is transfered through the hybridization to the impurity density of states 
and small steps are observed 
for $\omega= D_{s}$. The peak of larger spectral weight near $\omega=E_d$ is the charge-transfer
peak. Since for $\omega \sim E_d$, there is no surface density of states, the total width 
at half maximum expected for this peak is $\sim 4 \Delta_b=2$,\cite{capa} in agreement with what we obtain.
This peak is almost unchanged as $D_s$ crosses the Fermi energy [$\tilde{E}_{d}(E_{d})$ increases a little bit,
see Eq. (\ref{edef})].

Instead, the width of the Kondo 
peak near the Fermi energy changes dramatically. This width is of the order of the Kondo
temperature $T_K$ and the absence of surface states for $D_s>0$ renders $T_K$
nearly three order of magnitude smaller. Mathematically this is caused by the absence of $\Delta_s$ 
in the exponent of Eq. (\ref{tkf}). In spite of this, the shape of the Kondo peak does not change too much.
Another difference apparent in the figure, is a factor near 2 between the intensity of the peak for
$D_s>0$ compared to that for $D_s<0$.
We remind the reader that due to the Friedel sun rule, the spectral density
at the Fermi energy can be written in the form \cite{lan,loig}

\begin{equation}
\rho _{d\sigma }(\epsilon _{F})=\frac{\sin ^{2}\varphi _{\sigma }}{\pi
\Delta _{\sigma }(\epsilon _{F})},  \label{l1}
\end{equation}
where

\begin{equation}
\varphi _{\sigma }=\pi \langle d_{\sigma }^{\dagger }d_{\sigma }\rangle +
\mathrm{Im}\int_{-\infty }^{\epsilon _{F}}d\omega G_{d\sigma }(\omega
+i\epsilon )\frac{\partial \Sigma _{0\sigma }(\omega +i\epsilon )}{\partial
\omega }.  \label{fsr}
\end{equation}%
In the usual case of a flat wide symmetric conduction band, $\partial \Sigma
_{0\sigma }/\partial \omega =0$ and the integral in Eq. (\ref{fsr}) can be
neglected. This is not our case. However we expect that the influence of
this term is rather small except when $\epsilon _{F}\sim D_{s}$. Taking into
account that the NCA has some deviations of the order of 10\% in Friedel sum
rule,\cite{su42} we do not calculate the integral here. Since for 
both $D_s$ we obtain an occupancy $0.47 < \langle d_{\sigma }^{\dagger }d_{\sigma }\rangle < 0.5$,
one expects $\varphi \sim \pi/2$ and $\rho_{d\sigma }(\epsilon _{F})$ slightly below 
$1/[\pi (\Delta_b + \Delta_s)] \approx 0.318$ for $D_s<0$ and 
$1/(\pi \Delta_b ) \approx 0.637$ for $D_s>0$. The corresponding NCA values are 0.315 and 0.614
respectively. In any case NCA tends to overestimate $\rho_{d\sigma }(\epsilon _{F})$.\cite{su42}

\begin{figure}[h]
\includegraphics[clip,width=8cm]{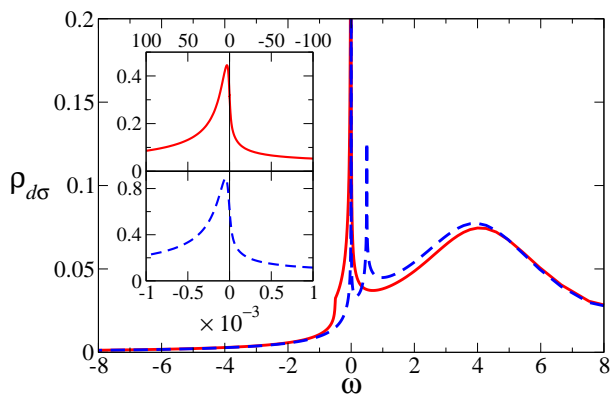}
\caption{(Color online) Same as Fig. \ref{electro} for $E_d+U=4$.}
\label{huecos}
\end{figure}

In \ref{huecos} we show the spectral density for the case in which the on site energies of the
impurity $E_d$ and $E_d+U$ are reflected through the Fermi energy, keeping $U \rightarrow \infty$.
This was calculated with NCA using an electron-hole transformation as explained in the previous section.
The charge transfer peak is now located for energies above $\epsilon_F$ and has a width near
$4(\Delta_s+\Delta_b)$, wider than in the previous case, 
because the surface states also contribute to its width. The Kondo temperatures
are larger than in the previous case, because now the shift given by the second member of Eq. (\ref{edef})
pushes the effective $d$ level towards the Fermi energy and then also $J$ increases [see Eq. (\ref{j})].
The structures for $\omega \sim D_s$ are more pronounced in this case, in particular for $D_s=0.5$, where
one can see a pronounced peak mounted on the left side of the charge-transfer peak and probably taken some
spectral weight from it. We have verified that in contrast to the Kondo peak, which as it is well known rapidly 
loses intensity with increasing temperature, the peak at $\omega = D_s$ for $D_s=0.5$ is practically independent 
of temperature for $T < T_K$.

Concerning the magnitude of 
the spectral density at the Fermi energy, we obtain with the NCA 
$\rho_{d\sigma }(\epsilon _{F})=0.315$ for $D_s <0$ and 0.597 for $D_s >0$, near to the maximum possible values
according to the Friedel sum rule. They are likely overestimated by a few \%.

\subsection{Numerical renormalization group}
\label{nrgr}

Here we present our NRG results for the same case and parameters shown in Fig. \ref{p5p5}.
We have determined $T_K$ in the same way as with the NCA, namely the temperature at which the conductance 
through the systems falls to half the ideal value.
To calculate the conductance we have calculated the Green function in each iteration which correspond to a 
temperature scale $\sim \lambda^{-0.5(N-1/2)}$ where $\lambda$ is the renormalization parameter and $N$ is 
the number of iteration, and we have used the z-trick \cite{Yoshida} to reduce the errors due 
to discretization.

\begin{figure}[h]
\includegraphics[clip,width=8cm]{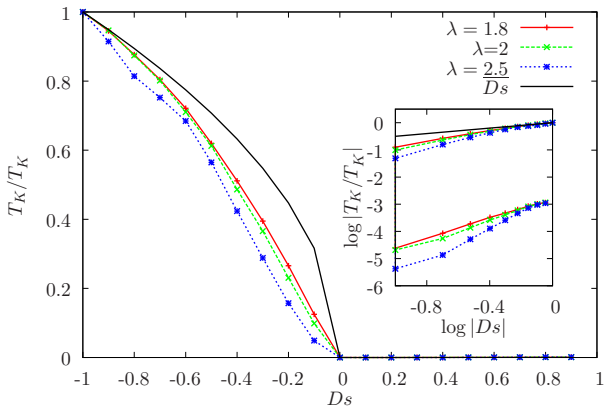}
\caption{(Color online) Kondo temperature as a function of the bottom of
the conduction band calculated with NRG for several values of $\Lambda$. 
The inset shows the results in log-log scale.
Parameters as in Fig. \ref{p5p5}. $T_{K}^{0}$ is the Kondo temperature 
for $D_s=-D$.}
\label{nrgf}
\end{figure}

The result is shown in Fig. \ref{nrgf} for several values of $\Lambda$. We see that for the largest value 
$\Lambda=2.5$ one observes some oscillations, suggesting that the algorithm has some difficulties in representing a 
step at finite energies due to the logarithmic discretization. As $\Lambda$ decreases, the curve 
for negative $D_s$ approaches the dependence with $D_s$ expected from PMS and NCA. 
Unfortunately, decreasing $\Lambda$ further would require a precision that is beyond our capabilities. 
Concerning the magnitude of the Kondo temperature, the NRG value for $D_s=-D$ is $T_{K}^{0}=0.0048$, 38 \% smaller 
than the corresponding NCA value 0.0077.
For $D_s>0$ the behavior of $T_K$ is also similar to the NCA result.

\section{Summary and discussion}

\label{sum}

We have calculated the dependence of the characteristic energy scale of the Kondo effect
$T_K$ for the impurity Anderson model in the presence of a step in the conduction
density of states.
This is the physical situation that takes place at the (111) surface of 
Cu, Ag and Au, where a two-dimensional band of surface Shockley states start 
at an energy $D_s$ slightly below the Fermi level $\epsilon _F$.
Depending on the element, $\epsilon _F-D_s$ ranges from 67 to 475 meV.
This difference can be changed by alloying the different noble
metals at the surface,\cite{cerce,abd} or by applying strain, changing the sign
of it,\cite{neu,andy1} as explained in Section \ref{intro}.

We obtain that in the general case, as $D_s$ is varied, $T_K$ can be well described by a power 
law $T_K \simeq |D_s-\epsilon_F|^{\eta}$ for $D_s-\epsilon_F<0$ and a different power law
$T_K \simeq (D_s-\epsilon_F)^{\zeta}$ for $D_s-\epsilon_F > 0$. 
This dependence is no more valid for very small $|D_s| \sim T_K$.
The Kondo temperature is much larger for negative 
$D_s-\epsilon _F$ because of the presence of surface states at the Fermi energy. 
The exponent is in general non-trivial, except 
for $U \rightarrow \infty$ and $D_s-\epsilon _F > 0$ ($<0$) if $E_d$ ($E_d+U$) is finite,
in which case $\zeta=0$ ($\eta=0$) and $T_K$ slightly increases with $D_s$. 
The exponents are given by Eq. (\ref{tkf}) and depend on the ratio 
between surface and bulk hybridizations
with the impurity $\Delta_s/\Delta_b$ and the ratio between on-site and
Coulomb energy $E_d/U$. Thus, changing $D_s-\epsilon_F$ by alloying or by other
method, might provide a way to extract these ratios 
which are difficult to estimate by alternative methods.

For $|D_s-\epsilon _F| < \Delta_s + \Delta_b$ and $\Delta_s \gtrsim \Delta_b$,
the spectral density of states of the magnetic impurity also shows steps
or peaks at $\omega \sim D_s$.

We expect that our work stimulates further experimental work on the subject.

We now discuss several effects that might be present in real systems absent in our model.
We have assumed a constant hybridization between the magnetic impurity 
added at the surface and both bulk and surface states. 
Actually, the important fact is that the  hybridization is rather featureless
in an energy range larger than   $2|D_s|$ around the Fermi energy. In general, one expects that this should be true for 
sufficiently small $|D_s|$, if symmetry allows it.
Some calculations suggest rather constant hybridization in a range of 1 eV around the Fermi energy.\cite{merino}

We have also assumed a non-degenerate magnetic orbital  
hybridizing with bulk and surface states
of the same symmetry (like a $d_{3z^2-r^ 2}$ localized state hybridizing with bulk and suface $s$ and $p_z$ states), 
leading to the simplest Anderson model to describe the system.
In some cases degenerate orbitals are expected. For example, for iron(II) phtalocyanine (FePc) molecules on Au(111),
the important orbitals are the degenerate Fe $3d$ Fe orbitals with symmetry $xz$ and $yz$ 
and the effective low-energy impurity model has SU(4) symmetry.\cite{mina,mole,joa2}
Our results can be easily extended for this model and similar power-law dependences would result.
However, in this case we expect that the hybridization vanishes at the bottom of the
surface band for symmetry reasons. In the hypothetical case of two electrons occupying both orbitals,
one has the two-channel spin-1 Kondo model, which is also a Fermi liquid \cite{dinap} and we expect
a similar physics.

One might wonder if Rashba spin-orbit coupling, which splits the Fermi wave vector of the surface states
of Au(111) in two values (0.160 \AA$^{-1}$ and 0.186 \AA$^{-1}$ \cite{soc}) affects our conclusions.
However, NRG calculations show that the effect on the Kondo temperature is very small.\cite{tksoc1,tksoc2}
Therefore, except perhaps for very small $D_s-\epsilon_F$, we do not expect a significant effect.

Finally, to estimate the effect of a non sharp edge in the surface spectral density of states,
we have calculated using PMS the Kondo temperature for a linear increase
of $\rho _{s}$ between $D_{s}-\delta $ and $D_{s}$ with $\delta >0$ for $D_{s}<0$. 
To linear order in $\delta /|D_{s}|$ the result is

\begin{equation*}
T_{K}(\delta )=T_{K}(0)\left[ 1+\frac{\Delta _{s}}{4(\Delta _{s}+\Delta _{b})%
}\frac{\delta }{|D_{s}|}\right] 
\end{equation*}%
Therefore, the correction is small except when $D_{s}$ is very near the
Fermi level.

Concerning the different techniques used, we obtain a very good agreement
between poor man's scaling (PMS) on the effective Kondo model and the non-crossing
approximation (NCA). Taking into account the success of both approaches in
similar problems, this is a further indication that these approximations
give accurate results for the Kondo temperature, apart from a factor of
the order of one in the smallest non-trivial order order in PMS.

The  numerical-renormalization group which is usually a very accurate 
technique 
for low-energy features has trouble in capturing the step in the conduction band,
due to the logarithmic discretization of the latter. 

The slave bosons in the mean-field
approximation (SBMFA) gives wrong exponents for the dependence of $T_K$ on $D_s-\epsilon_F$.
It seems to miss the renormalization of the effective exchange constant.
In spite of this for not too small $|D_s-\epsilon_F|$ it gives the correct order of magnitude of
$T_K$. Usually $|D_s-\epsilon_F| \gg T_K$. In one of the worst cases, 
Co on Ag(111),  the ratio $T_K/(\epsilon_F- D_s) $ is slightly below 0.1.\cite{serra}
For this ratio and some cases we have studied, the SBMFA overestimates $T_K$
by a factor near 4 in comparison with NCA [in addition to the prefactor $A$ 
in Eq. (\ref{tkf})], while it works better for small $\Delta_s$ or $E_d+U$
near the Fermi energy. 

These results are useful for researchers studying similar problems.

\section*{Acknowledgments}

We thank Pablo Cornaglia for useful discussions and assistance in 
the NRG calculations. This work was sponsored by PIP 112-201101-00832 of CONICET and PICT
2013-1045 of the ANPCyT.

\end{document}